\documentclass[journal]{IEEEtran}
\usepackage{graphicx,subfigure}
\usepackage{url}
\usepackage[]{amsmath}
\usepackage{amssymb}

\newcommand{\R}{\mathbb{R}}
\usepackage{url}
\graphicspath{{Images/}}
\usepackage{xcolor}
\usepackage{amsmath} 
\usepackage{float}
\usepackage[numbers]{natbib}

\begin{document}
\title{SpectralNET: Exploring Spatial-Spectral WaveletCNN for Hyperspectral Image Classification}

\author{Tanmay Chakraborty and Utkarsh Trehan\\
MALIS - Machine Learning and Intelligent Systems\\
Lecturer: Dr. Maria A. Zuluaga\\
Department of Data Science and Engineering, EURECOM, Biot, France.\\
Emails: tanmay.chakraborty@eurecom.fr, utkarsh.trehan@eurecom.fr
}

\maketitle

\begin{abstract}
 Hyperspectral Image (HSI) classification using Convolutional Neural Networks (CNN) is widely found in the current literature. Approaches vary from using SVMs to 2D CNNs, 3D CNNs, 3D-2D CNNs. Besides 3D-2D CNNs and FuSENet, the other approaches do not consider both the spectral and spatial features together for HSI classification task, thereby resulting in poor performances. 3D CNNs are computationally heavy and are not widely used, while 2D CNNs do not consider multi-resolution processing of images, and only limits itself to the spatial features. Even though 3D-2D CNNs try to model the spectral and spatial features their performance seems limited when applied over multiple dataset. In this article, we propose SpectralNET, a wavelet CNN, which is a variation of 2D CNN for multi-resolution HSI classification. A wavelet CNN uses layers of wavelet transform to bring out spectral features. Computing a wavelet transform is lighter than computing 3D CNN. The spectral features extracted are then connected to the 2D CNN which bring out the spatial features, thereby creating a spatial-spectral feature vector for classification. Overall a better model is achieved that can classify multi-resolution HSI data with high accuracy. Experiments performed with SpectralNET on benchmark dataset, i.e. Indian Pines, University of Pavia, and Salinas Scenes confirm the superiority of proposed SpectralNET with respect to the state-of-the-art methods. The code is publicly available in https://github.com/tanmay-ty/SpectralNET.

\end{abstract}


\IEEEpeerreviewmaketitle

\section{Introduction}
\IEEEPARstart{A}{} Hyperspectral Image (HSI) is a high dimension image cube, where each band stores the intensity values of the pixels in a particular spectrum \cite{AMIGO20203}. HSI classification is the task of correctly predicting the different pixel values associated with the different classes present in a remotely sensed HSI. Applications include urban development, detection of land changes, military applications, land cover analysis, crop detection etc. A key feature of HSI is they contain both spectral and spatial information.

Deep-learning based methods specially CNNs perform extremely well on image data. In recent works, HSI classification using different CNN models is also seen besides traditional hand-extracted feature based models \cite{8697135}. Most models are based on 2D CNN, and 3D CNN \cite{8340197}. Due to satisfactory performances of the two independent models \cite{9200676}, hybrid 3D-2D CNNs have also been proposed in the literature \cite{HAN202038}. FuSENet is another model proposed in literature for HSI classification \cite{iet:/content/journals/10.1049/iet-ipr.2019.1462}.

In \cite{7465721}, a band weighing strategy has been proposed that utilizes multiple binary support vector machines (SVM) in order to maximize the spectral distances between each class of a remotely sensed HSI. Their method was able to weight the spectral bands and improve classification results. A similar approach using SVMs has been proposed in \cite{7831354}, where the authors explored discrete space model (DSM) to transform continuous spectral features into discrete feature space, they utilized a composite kernel to take into account the spectral and spatial features. This pre-processing step improved the performance of SVMs for HSI classification. Kernel based approaches has also been found in the literature. In \cite{wang2020spectral}, spectral similarity based kernels has been developed and utilized along with the RBF kernel in a SVM. For the problem in hand they concluded spectral similarity based kernels outperform traditional SVM kernels. 

The work in \cite{8447427}, adapts and improves the traditional low-rank representation (LRR) to the HSI classification problem. Locality-and structure-regularized LRR combines both the spectral and spatial features to explore the local similarity of pixels. The authors of \cite{CHUNHUI201861}, applied the concept of spectral gradient for HSI classification. They extracted the spatial features through a random forest algorithm and spectral features through spectral gradients. Then they perform a multi-scale fusion to integrate spatial-spectral features for the SVM to perform classification. The work in \cite{OKWUASHI2020107298}, introduced deep support vector machines (DSVM) for HSI classification. The model was able to outperform most of the state-of-the-art algorithms including all the variants of traditional SVMs. %the authors of \cite{9050922}, focused on extracting spatial-spectral deep features and fused them together to create a feature vector, which was then utilized for classification. They suggested using the densely connected architectures to deal with the vanishing gradient problem. 

In \cite{xu2020csa}, a 3D octave CNN has been proposed which factorizes the mixed frequency feature map to reduce the spatial redundancy obtained when using a traditional 3D CNN with HSI. The authors of \cite{8851917}, utilized pseudo 3D blocks with a densely connected network. Their pseudo 3D blocks can capture both spectral and spatial features simultaneously compared to a traditional 3D CNN. The article \cite{ahmad2020fast}, utilized small 3D patches extracted from the original HSI cube to train a 3D CNN with 3D kernel. In the following works \cite{8061020}, residual connections were added to a 3D CNN in order to assimilate both high and low level features present in a HSI and improve classification results. The work of \cite{10.1007/978-3-030-14118-9_2}, studied the effect of dimensionality reduction of HSI on 3D CNNs. They concluded reducing the dimension of the training image reduced training time by 60\%. 

In \cite{9103280}, a 3D-2D CNN has been proposed for HSI classification. As a pre-processing the authors utilized channel wise shift and channel wise weighting to highlight the different spectral bands. In \cite{9200676}, 2D-3D CNN has been utilized with multi band feature fusion mechanism. This mechanism allows them to fuse both shallow and deep features in spectral band, which improves the feature vector sent into the dense layer.% A rotation equivariant CNN has been proposed in \cite{9208732}, they utilized a CNN with circular harmonic filters to automatically tackle the orientation problem creeping in from the image source. 
The work proposed in \cite{fang2020combining}, introduces adaptive spectral unmixing into a 3D-2D CNN along with a early exit strategy. The early exit strategy reduces computational cost for easy samples. In \cite{feng2019learning}, a residual hybrid 3D-2D CNN has been proposed, which has further been improved in \cite{8736016} and is currently the state-of-the-art.

Efforts have also been made with Recurrent Neural Networks (RNN), Generative Adversarial Networks (GAN), Graph CNNs \cite{9091940}, and Squeeze and Excitation Residual Network \cite{wang2019spatial}. RNNs consider the spectral signature of the HSI as a sequence in order to learn discriminative features \cite{8662780}. 

Even though the 3D-2D CNNs model both the spatial and spectral features from a HSI cube, their model performance when applied over multiple dataset seems limited. 3D CNNs are also computationally expensive over 2D CNNs. So a method involving only 2D CNN as well as the power of extracting both spatial and spectral features is desirable. 

In this article, a 2D wavelet CNN has been proposed for HSI classification. The work in \cite{PRABHAKAR201737}, established wavelet transform as a good feature extractor for HSI classification task. Thus fusing the wavelet transform into a 2D CNN model brings out both the spectral and spatial features from a HSI. These features are then concatenated channel wise and sent as an input to the dense classification layers of the 2D CNN. The developed model uses Factor Analysis (FA) as a pre-processing step to reduce the huge dimensionality of HSI. Then patches are extracted and sent into the CNN. This reduces the training time as well. The spectral features coming from wavelet transform are computationally lighter as well compared to a 3D CNN. The model outperforms all previous models and paves the way for wavelet CNN in multi-resolution image classification. This model has been named \textit{SpectralNET} in this paper.  

The rest of the paper is arranged in the following way Section II, describes the \textit{SpectralNET} model in details, Section III contains our experiments and discussions, and the paper is concluded in Section IV.

\section{SpectralNET}

The conventional 2D CNN can be considered a limited version of a multi-resolution CNN that can consider both spectral and spatial information \cite{fujieda2018wavelet}. Previous works have been successful in establishing the convolution and pooling function in a 2D CNN as filtering and downsampling \cite{fujieda2017wavelet}. A basic CNN can be mathematically represented as the weighed sum of nearest neighbours with an added constant bias. 

\subsection{Background for SpectralNET}

\begin{figure*}[!tb]
    \centering
    \includegraphics[scale=0.18]{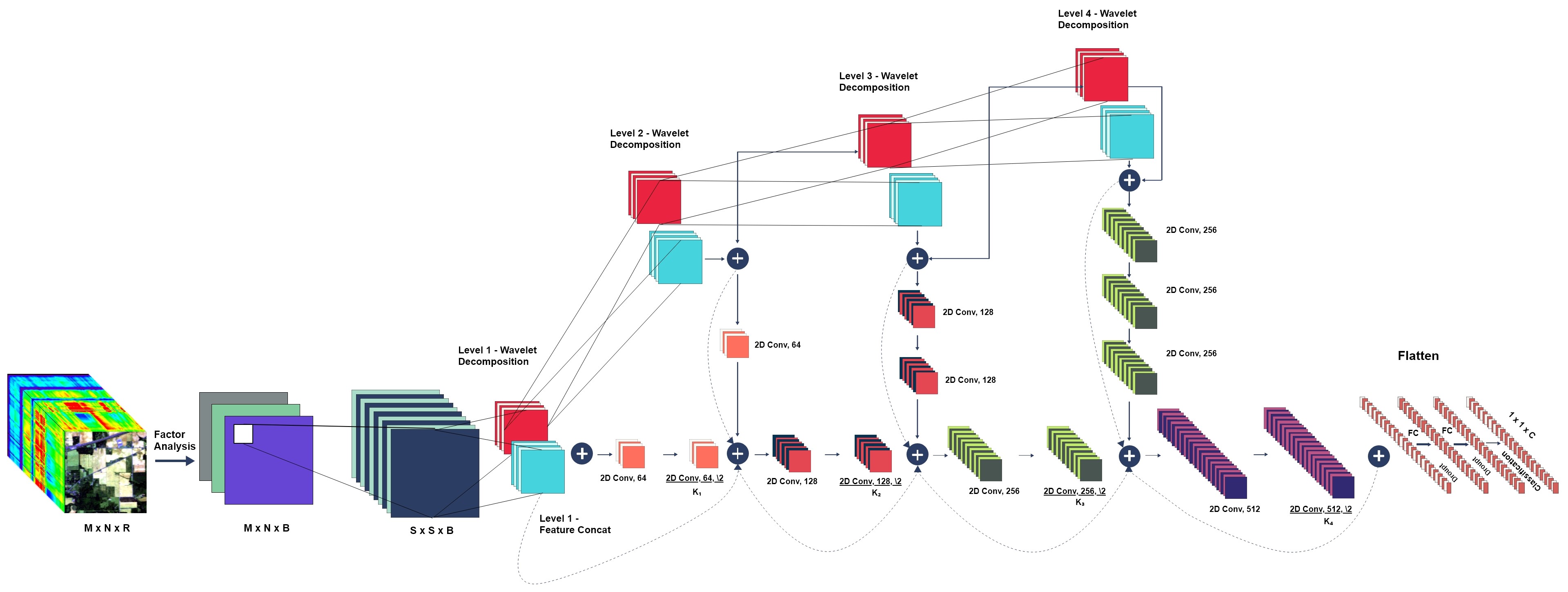}
    \caption{Input HSI cube is pre-processed using Factor Analysis (FA) to reduce the dimention to 3. Patches are extracted from the pre-processed image and sent as an input to the SpectralNET model. SpectralNET model architecture with 4-level wavelet decomposition of the input HSI patch. The input kernel size is 3x3 with 1x1 padding. The output batch channel size is denoted by the numbers written after \textit{conv}. To reduce feature map 3x3 kernels with stride 2 and 1x1 padding are used. The wavelet transformed features are added channelwise. To prevent the gradient from vanishing projection shortcuts are utilized with 1x1 convolutions. An average pooling layer is used globally after which the output is sent to the fully connected layers with dropout neurons.}
    \label{fig:architecture}
\end{figure*}

Given an input vector $X_n$ with corresponding labels $Y_n$ from the $\R^n$ space. In equation \ref{eq1}, $Y_i$ is a label from $\mathit{Y_n}$ labels and $X_i$ is the corresponding sample from $\mathit{X_n}$. $W_j$ is the weight defined by a filtering kernel. $N_i$ are neighbouring \textit{i} data.   

\begin{equation}
    y_i =  \sum_{j\in N_i} W_j X_j 
    \label{eq1}
\end{equation}

The equation \ref{eq1} can be simply considered as the convolution of $X_i$ and kernel $W_j$ and can be rewritten as $\mathbf{Y = X*W}$. This is called the convolution layer of a CNN, where $\mathbf{W}$ is in $\R^o$. The output of the convolution layers are typically big and needs to be pooled down before feeding it to the next layer. The pooling layers are placed in between convolution layers to perform a filtering operation and reducing the number of outputs. 

%The mean pooling function performs an average based on a parameter $\mathbf{P}$, the mean filter, essentially it indicates the scale of reduction of the input. The pooling operation is given in equation \ref{eq2}, where $\mathbf{X_O}$ in $\R^c$ is the input and $\mathbf{y_p}$ in $\R^a$ is the mean pooled output $a = \frac{c}{P}$. The equation combining convolution and pooling can be written as $\mathbf{y} = (\mathbf{X} * \mathbf{k}) \downarrow \textit{p}$. The weight $\mathbf{k}$ when $\mathit{P} > 1$ is $\mathbf{W*P}$.  

%\begin{equation}
%    y_{P{_j}} = \frac{1}{P} \sum_{k = 0}^{P-1} X_{O_{pj}} + k 
%    \label{eq2}
%\end{equation}

This paves the way towards the multi-resolution CNN where the convolution is performed by a pair of kernels $\mathbf{k_{low}}$ and $\mathbf{k_{high}}$ which generate $\mathbf{X_{low}}$ and $\mathbf{X_{high}}$. The multi-resolution CNN performs the hierarchical decomposition of the $\mathbf{X_{low,t}}$ into $\mathbf{X_{low,t+1}}$ and $\mathbf{X_{high,t+1}}$ with different kernels at each step \textit{t}. 

For \textit{SpectralNET}, the wavelet kernel $\mathbf{K_{high,t}}$ is Haar wavelets and $\mathbf{K_{low,t}}$ is a scaling function  \cite{wang2006moving}. The 2D haar wavelets utilize the following four kernels ($f_{L,L} f_{L,H} f_{H,L} f_{H,H}$) for wavelet transform \cite{liu2018multi}. 
\begin{equation}
\begin{split}
f_{L,L} =
\begin{bmatrix}
1&1\\1&1
\end{bmatrix}
f_{L,H} =
\begin{bmatrix}
-1&-1\\1&1
\end{bmatrix}
\\
f_{H,L} =
\begin{bmatrix}
-1&1\\-1&1
\end{bmatrix}
f_{H,H} =
\begin{bmatrix}
1&-1\\-1&1
\end{bmatrix}
\end{split}
\end{equation}

A HSI patch \textbf{x} with SxS dimensions when passed through a Haar transform the \textit{(i,j)}-th spectrum position value can be written as $Haar(\mathit{i,j}) = x(2i-1, 2j-1) + x(2i-1, 2j) + x(2i, 2j-1) + x(2i, 2j)$.

The HSI patch taken as an input is decomposed by the wavelet transform into sub-bands, these sub-bands are then sent through a convolution layer to learn the spectral and location features. Note that the sub-bands indicated as high and low pass filers do not necessarily filter the spectral band in with high pass and low pass filter. The part of the sub-band is again decomposed in the next layer by the wavelet transform and sent into the convolution layer. This process is continued in each layer and the CNN continues to learn the spectral and spatial features from the HSI patch.

\subsection{SpectralNET Model Description}
The input HSI cube having dimension \textbf{MxNxR} is first sent into a layer of Factor Analysis (FA) to reduce the dimension into \textbf{MxNxB}. Reducing the dimension reduces training time by 60\% \cite{10.1007/978-3-030-14118-9_2}. The output vector \textbf{Y} having a dimension \textbf{1xMN} take up a class from the available land cover categories denoted by \textbf{C}. The spectral dimensions are preserved in FA, i.e. \textbf{MxN}, just the bands are reduced from \textbf{R} to \textbf{B}. Using FA in HSI as a pre-processing step is extremely beneficial, as FA is able to describe the variability among the different correlated and overlapping spectrum bands, which helps making the model classify similar examples better. On the other hand, commonly used Principal Component Analysis (PCA) based reduction does not directly address this objective in HSI. PCA provides an approximation to the required factors which do not help to differentiate similar examples that well. After the FA step is complete, overlapping 3D patches of size \textbf{SxSxB} are extracted from the pre-processed HSI and sent into the SpectralNET. \textbf{SxS} is the window size for patch extraction, for the Indian Pines dataset the patch size has been set at \textbf{64x64} and for the University of Pavia and Salinas Scene dataset the window size has been set at \textbf{24x24}. The truth values for these patches are determined by the center pixel's class category. The values were chosen based on experimentation to maximize the overall accuracy. 

\subsection{Implementation}
The proposed \textit{SpectralNET} model architecture is given in figure \ref{fig:architecture}. The model is initialized with 3x3 convolution kernels and 1x1 padding. To replace pooling layers in between convolution a stride of 2 has been utilized. A global mean pooling has been employed at the end of all the convolution layers before sending into the dense layer, this prevents overfitting in the model. Dense connections has been utilized along with projection shortcuts for utilizing the wavelet transformed data more efficiently \cite{he2016deep} \cite{huang2017densely}. Dense connections with channel wise concatenation of the decomposed data makes sure that all the features flow till the end of the model. The model explored two dropout layers as well along with batch normalization to prevent overfitting. Since the number of samples are very less in HSI the chances of overfitting are high. All steps to prevent the model from overfitting needs to be taken. Rectified Linear Unit (ReLU) has been utilized as the activation function. We explored the Stochastic gradient descent (SGD) over 150 epochs with a learning rate of 0.01 and momentum of 0.9 to optimize the objective function. %The trainable parameters are given in table \ref{params}.

%\begin{table}[]
%    \centering
%    \caption{Parameters of the model}
%    \begin{tabular}{ c  c }
%        \hline
%         Non-trainable params & 4736\\
%         
%         Trainable params & 6,800,336\\
%         \hline
%         Total params & 6,805,072\\
%         \hline
%    \end{tabular}
%    \label{params}
%\end{table}

\section{Experiments, results, and discussion}

\begin{figure*}[]
\centering
\subfigure[IP Image cube]{\includegraphics[height = 2.5cm, width = 2.5cm]{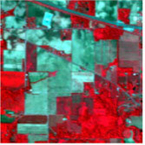}}\quad
\subfigure[IP Ground Truth]{\includegraphics[height = 2.5cm, width = 2.5cm]{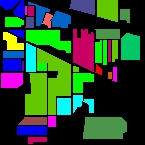}}\quad
\subfigure[IP Predictions]{\includegraphics[height = 2.5cm, width = 2.5cm]{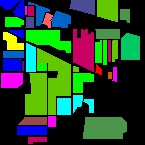}}\quad
\subfigure[IP legend]{\includegraphics[height = 2.5cm, width = 4.5cm]{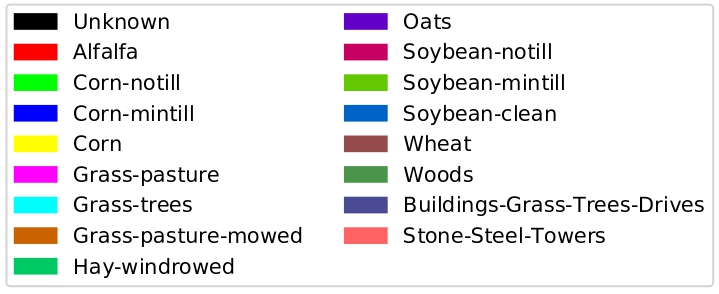}}\quad
\caption{Image cube, spectral ground truth, and spectral prediction for Indian Pines Dataset along with legend.}
\label{fig:ipdat}
\end{figure*}

\subsection{Dataset and Training}
The experiments were conducted on multiple publicly available benchmark datasets, Indian Pines (IP), University of Pavia (UP), and Salinas Scene (SA) \footnote{http://lesun.weebly.com/hyperspectral-data-set.html}. The detailed descriptions of the three datasets are given in table \ref{tab:dataset}.The classification spectral layout for IP dataset is given in figure \ref{fig:ipdat}.

To perform the experiments, Google colab cloud platform with GPU has been utilized\footnote{https://colab.research.google.com/}. Based on our experimental analysis an optimum learning rate of 0.01 with a momentum of 0.9 was chosen for the SGD optimizer. For preserving the validity of the results for all datasets, the bands of the extracted patches have all been set to 3. So, the patch dimension for IP dataset is 64x64x3 and for UP and SA it is 24x24x3 respectively. The model has been trained for 150 epochs and convergence was achieved at around 60 epochs.
%The model has been trained for 150 epochs and the loss convergence curve given in figure \ref{fig:loss} shows that the model converged fast, at around 60 epochs. Which indicates the proposed model is efficient in HSI classification task. 

\begin{table}[]
    \centering
    \caption{Detailed description of each dataset used during experiment.}
    \begin{tabular}{c c c c c}
         Name & Spatial Dimension & Spectral Bands & Wavelength Range & Classes \\
         \hline
         IP & 145x145 & 224 & 400nm - 2500nm & 16 \\
         UP & 610x340 & 103 & 430nm - 860nm & 9 \\
         SA & 512x217 & 224 & 360nm - 2500nm & 16\\
         \hline
    \end{tabular}
    \label{tab:dataset}
\end{table}
\subsection{Classification Results}
The classification results are given in table \ref{tab:accuracy}. Three benchmark metrics are utilized to judge the performance of the proposed model. Overall Accuracy (OA) gives the total number of correctly classified labels out of the total number of labels. Average Accuracy (AA) gives the mean of class wise classification accuracies, and Kappa Accuracy is a measure that correlates the ground truth and classified values. The results are compared with the state-of-the-art methods like HybridSN \cite{8736016} and FuSENET \cite{iet:/content/journals/10.1049/iet-ipr.2019.1462}, besides SVM, 2D CNN, 3D CNN, M3D CNN \cite{8297014} \footnote{https://github.com/eecn/Hyperspectral-Classification}. The results are compared for two sets 10\% - 90\% random train test split and 30\% - 70\% random train test split respectively.

It can be observed from the results that the proposed model outperforms all state-of-the-art models in both the sets. Even though in the 10\% train set the HybridSN model appears to perform better in SA dataset, that might be because of the fact it takes a lot more spectral bands as input compared to the proposed model. It can also be seen from the results that 2D CNN standalone performs better than 3D CNN in SA dataset. It might be due to the increased spectral redundancy in the SA dataset compared to the rest. The performance of FuSENET, HybridSN and \textit{SpectralNET} is consistently high throughout the three dataset over M3D CNN. \textit{SpectralNET} is able to outperform all even with a lot less spectral bands, i.e. 3, utilized than the state-of-the-art models which utilize 15, 30 bands. This highlights the merit of using wavelets based spectral features with a CNN. The time for training the \textit{SpectralNET} is around 30 minutes which is also comparable to the currently established models. 

For more detailed class wise classification results are in the appendix. From the results it can be established that the performance of \textit{SpectralNET} is superior to all the methods currently available for HSI classification. 

%\begin{figure}
%    \centering
%    \includegraphics[scale = 0.35]{Images/losscurve.PNG}
%    \caption{Loss convergence curve for the UP dataset.}
%    \label{fig:loss}
%\end{figure}

\begin{table*}
    \centering
    \caption{Classification accuracies (\%) of proposed SpectralNET in terms of OA, Kappa, and AA with varying training data 10\% and 30\%, respectively}
    \label{tab:accuracy}
    \resizebox{!}{2cm}{
    \begin{tabular}{c | c | c c c | c c c | c c c}
            
        Training Samples & Methods &  & IP dataset &  &  & UP dataset &  &  & SA dataset & \\
        &  & OA & Kappa & AA & OA & Kappa & AA & OA & Kappa & AA\\
        \hline
        
         & SVM & 81.67$\pm$0.65 & 78.76$\pm$0.77 & 79.84$\pm$3.37 & 90.58$\pm$0.47 & 87.21$\pm$0.70 & 92.99$\pm$0.36 & 94.46$\pm$0.12 & 93.13$\pm$0.34 & 93.01$\pm$0.60 \\
         
         & 2D-CNN & 80.27$\pm$1.2 & 78.26$\pm$2.1 & 68.32$\pm$4.1 & 96.63$\pm$0.2 & 95.53$\pm$0.2 & 94.84$\pm$1.4 & 96.34$\pm$0.3 & 95.93$\pm$0.9 & 94.36$\pm$0.5 \\
         
         & 3D-CNN & 82.62$\pm$0.1 & 79.25$\pm$0.3 & 76.51$\pm$0.1 & 96.34$\pm$0.2 & 94.90$\pm$1.2 & 97.03$\pm$0.6 & 85.00$\pm$0.1 & 83.20$\pm$0.7 & 89.63$\pm$0.2 \\
         
        10\% & M3D-CNN & 81.39$\pm$2.6 & 81.20$\pm$2.0 & 75.22$\pm$0.7 & 95.95$\pm$0.6 & 93.40$\pm$0.4 & 97.52$\pm$1.0 & 94.20$\pm$0.8 & 93.61$\pm$0.3 & 96.66$\pm$0.5 \\
         
         & FuSENet & 97.11$\pm$0.2 & 97.25$\pm$0.2 & 97.32$\pm$0.2 & 97.65$\pm$0.3 & 97.69$\pm$0.3 & 97.68$\pm$0.4 & 99.23$\pm$0.1 & 99.97$\pm$0.2 & 99.16$\pm$0.1 \\
         
         & HybridSN & 98.39$\pm$0.1 & 98.16$\pm$0.1 & 98.01$\pm$0.2 & \textbf{99.72$\pm$0.1} & \textbf{99.64$\pm$0.1} & 99.20$\pm$0.1 & \textbf{99.98$\pm$0.2} & \textbf{99.98$\pm$0.2} & \textbf{99.98$\pm$0.1} \\
         
         & \textbf{SpectralNET} & \textbf{98.76$\pm$0.2} & \textbf{98.59$\pm$0.1} & \textbf{98.61$\pm$0.1} & 99.71$\pm$0.1 & 99.62$\pm$0.1 & \textbf{99.43$\pm$0.2} & 99.96$\pm$0.2 & 99.96$\pm$0.1 & 99.97$\pm$0.1 \\
         
         \hline
         \hline
        
         & SVM & 87.24$\pm$0.38 & 85.27$\pm$0.45 & 85.15$\pm$1.10 & 95.65$\pm$0.13 & 94.63$\pm$0.17 & 94.60$\pm$0.14 & 94.95$\pm$0.10 & 94.48$\pm$0.11 & 97.93$\pm$0.11 \\
         
         & 2D-CNN & 88.90$\pm$1.3 & 87.01$\pm$1.6 & 85.70$\pm$1.0 & 96.50$\pm$0.4 & 96.55$\pm$0.3 & 96.00$\pm$0.1 & 96.75$\pm$0.6 & 96.71$\pm$0.7 & 98.57$\pm$0.2 \\
         
         & 3D-CNN & 90.23$\pm$0.2 & 89.70$\pm$0.3 & 89.87$\pm$0.1 & 97.90$\pm$0.3 & 97.22$\pm$0.1 & 97.30$\pm$0.1 & 95.54$\pm$0.5 & 94.81$\pm$0.3 & 97.09$\pm$0.6 \\
         
        30\% & M3D-CNN & 95.67$\pm$0.1 & 94.70$\pm$0.3 & 94.60$\pm$0.6 & 97.60$\pm$0.2 & 96.50$\pm$0.6 & 98.00$\pm$0.1 & 94.99$\pm$0.3 & 95.40$\pm$0.1 & 96.28$\pm$0.2 \\
         
         & FuSENet & 99.01$\pm$0.2 & 98.60$\pm$0.1 & 98.64$\pm$0.1 & 99.42$\pm$0.2 & 99.21$\pm$0.3 & 99.33$\pm$0.2 & 99.68$\pm$0.2 & 99.74$\pm$0.1 & 99.69$\pm$0.1 \\
         
         & HybridSN & 99.75$\pm$0.1 & 99.71$\pm$0.1 & 99.63$\pm$0.2 & 99.98$\pm$0.1 & 99.98$\pm$0.2 & 99.97$\pm$0.2 & 100 & 100 & 100 \\
         
         & \textbf{SpectralNET} & \textbf{99.86$\pm$0.2} & \textbf{99.84$\pm$0.2} & \textbf{99.98$\pm$0.1} & \textbf{99.99$\pm$0.1} & \textbf{99.98$\pm$0.1} & \textbf{99.98$\pm$0.1} & \textbf{100} & \textbf{100} & \textbf{100}\\
         
         \hline
         \hline        
        \end{tabular}}
\end{table*}

\section{Conclusion}
In a nutshell, a wavelet CNN has been proposed in this work for HSI classification task. The developed SpectralNET takes into consideration both spectral and spatial features present in a high dimensional HSI cube using layers of wavelet decomposition of the input and adding that to the CNN. Experiments conducted with the three benchmark datasets IP, UP and SA along with a comparison with the state-of-the-art methods establish the superiority of the proposed model.

This work has been done in the context of the Machine Learning and Intelligent System (MALIS) course and it represents the final project report.

\bibliographystyle{ieeetr}
\bibliography{spectralnet}

\clearpage

\begin{figure}[t]
\centering
\subfigure[IP Confusion Matrix]{\includegraphics[height = 6cm, width = 6cm]{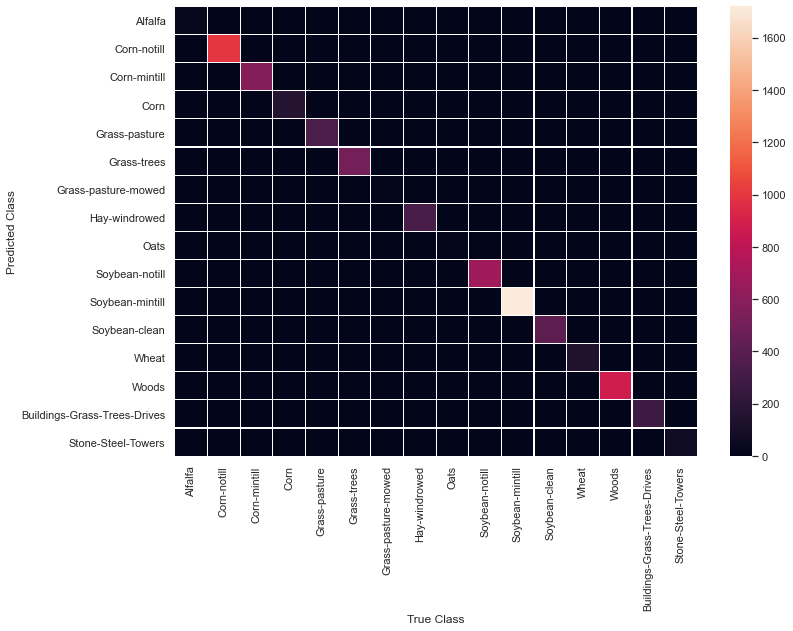}}\quad
\subfigure[UP Confusion Matrix]{\includegraphics[height = 6cm, width = 6cm]{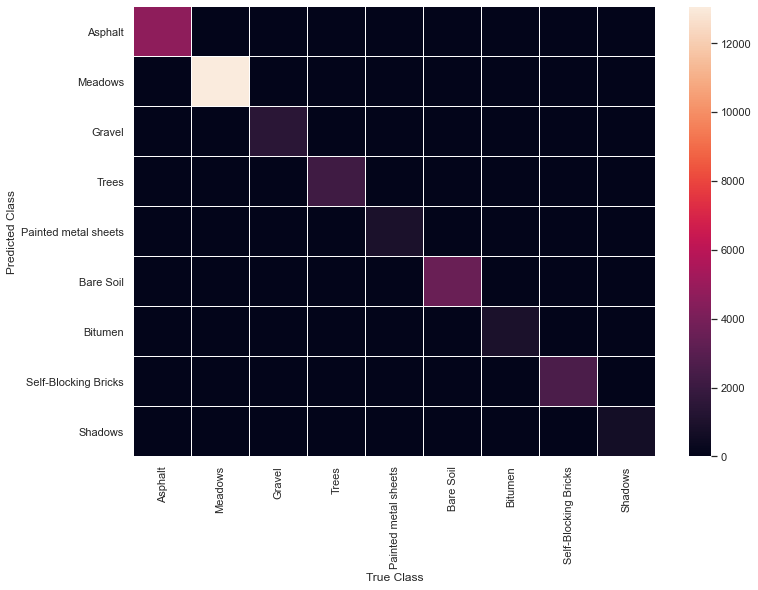}}\quad
\subfigure[SA Confusion Matrix]{\includegraphics[height = 6cm, width = 6cm]{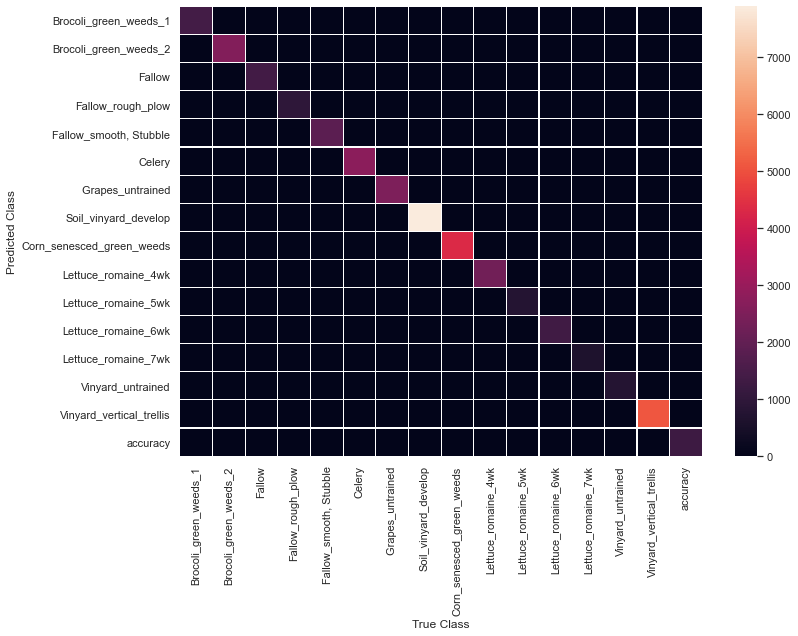}}\quad
\caption{Confusion matrix for IP, UP, and SA using SpectralNET.}
\label{fig:confu}
\end{figure}

\appendix[Classwise Classification Results]
Class wise classification results for IP, SA and UP datasets are summarised in table \ref{tab:ip_acc}, \ref{tab:sa_acc}, and \ref{tab:up_acc}  respectively. Confusion matrix are available in figure \ref{fig:confu}.

\begin{table}[htb]
    \centering
    \caption{Detailed classification results for Indian Pines Dataset in terms of Precision, Recall, f1-score, Test Loss, Overall Accuracy, Average Accuracy and Kappa Accuracy.}
    \resizebox{6cm}{!}{
    \begin{tabular}{c c c c c c}
         Class Labels & Precision & Recall & f1-score & Support\\
         \hline
         Alfalfa & 1.00 & 1.00 & 1.00 & 32\\
         Corn-notill & 1.00 & 1.00 & 1.00 & 1000\\
         Corn-mintill & 1.00 & 0.99 & 1.00 & 581\\
         Corn & 1.00 & 1.00 & 1.00 & 166\\
         Grass-pasture & 0.99 & 1.00 & 1.00 & 338\\
         Grass-trees & 1.00 & 1.00 & 1.00 & 511\\
         Grass-pasture-mowed & 1.00 & 0.85 & 0.92 & 20\\
         Hay-windrowed & 1.00 & 1.00 & 1.00 & 335\\
         Oats & 0.78 & 1.00 & 0.88 & 14\\
         Soyabean-notill & 1.00 & 1.00 & 1.00 & 680\\
         Soyabean-mintill & 1.00 & 1.00 & 1.00 & 1719\\
         Soyabean-clean & 1.00 & 1.00 & 1.00 & 415\\
         Wheat & 1.00 & 1.00 & 1.00 & 143\\
         Woods & 1.00 & 1.00 & 1.00 & 886\\
         Buildings-Grass-Trees-Drives & 1.00 & 1.00 & 1.00 & 270\\
         Stone-Steel-Towers & 0.98 & 1.00 & 0.99 & 65\\
         \hline
         accuracy &  &  & 1.00 & 7175\\
         macro avg & 0.98 & 0.99 & 0.99 & 7175\\
         weighted avg & 1.00 & 1.00 & 1.00 & 7175\\
         Test loss & & & & 0.7\% \\
         Average accuracy (\%) & & & & 99.98\% \\
         Kappa accuracy (\%) & & & & 99.84\% \\
         Overall accuracy (\%) & & & & 99.86\% \\
         \hline
    \end{tabular}}
    \label{tab:ip_acc}
\end{table}
\begin{table}[htb]
    \centering
    \caption{Detailed classification results for Salinas Scene Dataset in terms of Precision, Recall, f1-score, Test Loss, Overall Accuracy, Average Accuracy and Kappa Accuracy.}
    \resizebox{6cm}{!}{
    \begin{tabular}{c c c c c c}
         Class Labels & Precision & Recall & f1-score & Support\\
         \hline
         Brocoli-green-weeds-1 & 1.00 & 1.00 & 1.00 & 1406\\
         Brocoli-green-weeds-2 & 1.00 & 1.00 & 1.00 & 2608\\
         Fallow & 1.00 & 1.00 & 1.00 & 1383\\
         Fallow-rough-plow & 1.00 & 1.00 & 1.00 & 976\\
         Fallow-smooth & 1.00 & 1.00 & 1.00 & 1875\\
         Stubble & 1.00 & 1.00 & 1.00 & 2771\\
         Celery & 1.00 & 1.00 & 1.00 & 2505\\
         Grapes-untrained & 1.00 & 1.00 & 1.00 & 7890\\
         Soil-vinyard-develop & 1.00 & 1.00 & 1.00 & 4342\\
         Corn-senesced-green-weeds & 1.00 & 1.00 & 1.00 & 2295\\
         Lettuce-romaine-4wk & 1.00 & 1.00 & 1.00 & 748\\
         Lettuce-romaine-5wk & 1.00 & 1.00 & 1.00 & 1349\\
         Lettuce-romaine-6wk & 1.00 & 1.00 & 1.00 & 641\\
         Lettuce-romaine-7wk & 1.00 & 1.00 & 1.00 & 749\\
         Vinyard-untrained & 1.00 & 1.00 & 1.00 & 5088\\
         Vinyard-vertical-trellis & 1.00 & 1.00 & 1.00 & 1265\\
         \hline
         accuracy &  &  & 1.00 & 37891\\
         macro avg & 1.00 & 1.00 & 1.00 & 37891\\
         weighted avg & 1.00 & 1.00 & 1.00 & 37891\\
         Test loss & & & & 0.001\% \\
         Average accuracy (\%) & & & & 100\% \\
         Kappa accuracy (\%) & & & & 100\% \\
         Overall accuracy (\%) & & & & 100\% \\
         \hline
    \end{tabular}}
    \label{tab:sa_acc}
\end{table}

\begin{table}[!b]
    \centering
    \caption{Detailed classification results for University of Pavia Dataset in terms of Precision, Recall, f1-score, Test Loss, Overall Accuracy, Average Accuracy and Kappa Accuracy.}
    \resizebox{6cm}{!}{
    \begin{tabular}{c c c c c c}
         Class Labels & Precision & Recall & f1-score & Support\\
         \hline
         Asphalt & 1.00 & 1.00 & 1.00 & 4642\\
         Meadows & 1.00 & 1.00 & 1.00 & 13055\\
         Gravel & 1.00 & 1.00 & 1.00 & 1496\\
         Trees & 1.00 & 1.00 & 1.00 & 2145\\
         Painted metal sheet & 1.00 & 1.00 & 1.00 & 942\\
         Bare soil & 1.00 & 1.00 & 1.00 & 3520\\
         Bitumen & 1.00 & 1.00 & 1.00 & 931\\
         Self-Blocking Bricks & 1.00 & 1.00 & 1.00 & 2577\\
         Shadows & 1.00 & 1.00 & 1.00 & 663\\
         \hline
         accuracy &  &  & 1.00 & 29944\\
         macro avg & 1.00 & 1.00 & 1.00 & 29944\\
         weighted avg & 1.00 & 1.00 & 1.00 & 29944\\
         Test loss & & & & 0.07\% \\
         Average accuracy (\%) & & & & 99.98\% \\
         Kappa accuracy (\%) & & & & 99.98\% \\
         Overall accuracy (\%) & & & & 99.99\% \\
         \hline
    \end{tabular}}
    \label{tab:up_acc}
\end{table}

\end{document}